\documentclass[a4paper,12pt]{article}
\usepackage{cite}
\usepackage{epsfig}
\usepackage{graphicx}   
\usepackage{url}
\usepackage{float}
\usepackage{color}
\usepackage{amsmath}
\usepackage{amssymb}
\usepackage{mathrsfs}
\usepackage[numbers,sort&compress]{natbib}

\newlength{\dinwidth}
\newlength{\dinmargin}
\begin{document}
\title{ \bf  Scalar leptoquark effects in the lepton flavor violating exclusive $b\to s \ell^-_i\ell^+_j$ decays}
\author
{\small Jin-Huan Sheng$^{1,2}$,
~~Ru-Min Wang$^2$,\thanks{ruminwang@sina.com}
~~Ya-Dong Yang$^{1}$
\\
{\scriptsize {$^1$ \it  Central China Normal University,  Wuhan, Hubei 430079, P.R.China}}\\
{\scriptsize {$$ \it Institute of Particle Physics and Key Laboratory of Quark and Lepton Physics (MOE)}}\\
{\scriptsize {$^2$ \it College of Physics and Electronic Engineering, Xinyang Normal University, Xinyang, Henan 464000, P.R.China}}\\
}

\date{}
 \maketitle
\vspace{-0.4cm}

\begin{abstract}
Leptoquarks have been suggested to solve a variety of discrepancies between the expected and observed phenomenon.
In this paper,
we investigate the effects of scalar leptoquarks on the lepton flavor violating $B$ meson rare decays which involve the quark level transition
$b \to s \ell^-_i \ell^+_j(i \neq j)$.
The leptoquark parameter spaces are constrained by using the recently  measured upper limits on
 $\mathcal{B}(B_s^0 \to \ell^-_i \ell^+_j)$ and $\mathcal{B}(B \to K^{(*)}\ell^-_i \ell^+_j)$. Using such constrained leptoquark parameter spaces, some relevant physical quantities are  predicted
and we find that the constrained new physics parameters in the leptoquark model have very
 obvious effects on the relevant physical quantities.
With future measurements of observables in $B \to K^{(*)}\ell^-_i \ell^+_j$ decays at the LHCb, more and more differentiated from the other new physics explanations could be tested.
\\ \\
{\bf Key words:} leptoquark; lepton flavor violating; lepton number violating
\\
\noindent {\bf PACS Numbers: 13.20.He, 12.15.Mm,  14.80.Sv}   
\vspace{1.5cm}
\end{abstract}

\section{Introduction}

Study for rare decays of $B$ mesons induced by the flavor changing neutral current (FCNC) process $b \to s(d)$
plays a very important role to test the Standard Model (SM) and to provide crucial information in our search for New Physic (NP) beyond the SM.
The SM contributions to the rare $B$ meson decays, which involve FCNC process $b\to s(d)$, are absent
at the tree level due to the Glashow-Iliopoulos-Maiani (GIM) mechanism and occur via the one-loop level.
Recently, some  anomalies occur in the exclusive $b\to s\ell^-\ell^+$ decays, for examples,  the experimental measurements  of $\mathcal{B}(B \to K^{(*)} \mu^-\mu^+)/\mathcal{B}(B \to K^{(*)} e^-e^+)$ and the $P_5'$ angular parameter in $B^0\to K^{*0}\mu^+\mu^-$ decay
 deviate from their SM predictions  by $2-3\sigma$ \cite{Aaij:2017vbb,Hiller:2003js,Descotes-Genon:2013wba,Abdesselam:2016llu,Wehle:2016yoi}.
In addition, lepton flavor violating (LFV) decays of $\mu \to e \gamma$, $Z\to e\mu(e\tau,\mu\tau)$ and $h \to \mu \tau$
 have been searched in LEP1, International Linear Collider (ILC) and
 CMS \cite{Happacher:2017hrj,Nehrkorn:2017fyt,CMS:2017onh}.
Lepton flavor non-universality of  $\bar{b} \to \bar{s}\ell^- \ell^+$ decays implys that LFV processes may be seen in $B$ decays \cite{Glashow:2014iga,Becirevic:2015asa,Varzielas:2015iva,Hiller:2016kry,Guo:2017gxp,Alok:2017jaf,Alok:2017sui}.
The experimental observation for LFV decays will provide unambiguous signal for NP beyond the SM.

It is well known that leptoquarks (LQs) are color-triplet boson particle which can couple to a quark
and a lepton at the same time and can occur in various extensions of the SM \cite{Georgi:1974sy,Pati:1974yy}.
They can also have spin-1 (vector LQs) or spin-0 (scalar LQs).
Scalar LQs can exist at  TeV scale in
extended  technicolor models \cite{Schrempp:1984nj,Gripaios:2009dq} as well as  in  quark and lepton composite models \cite{lepto2}.
The phenomenology of scalar LQs have been studied extensively in many literatures
\cite{Dorsner:2009cu,Fajfer:2008tm,Povarov:2010vx,Saha:2010vw,Dorsner:2011ai,Kosnik:2012dj,
Queiroz:2014pra,Allanach:2015ria,Arnold:2013cva,Becirevic:2017jtw}.
It is generally assumed that the vector LQs tend to couple directly to neutrinos and hence expected
that their couplings are tightly constrained from the neutrino mass and mixing data. We  will only consider the model which a LQ can couple to a pair of quark and lepton and
may be inert with respect to  proton decay. Hence, the bounds from proton decay may not be applicable for such cases and
LQs may produce signatures in other low-energy phenomena \cite{Arnold:2013cva}.

In this paper, we will investigate the LFV exclusive $b \to s \ell^-_i \ell^+_j$ decays
in the scalar LQ model.
The upper limits on the relevant lepton number violating (LNV)
coupling products due to the scalar LQ exchanges
 are obtained from the recent limits of $\mathcal{B}(B_s \to \ell^-_i\ell^+_j)$ and $\mathcal{B}(B \to K^{(*)}  \ell^-_i\ell^+_j)$.
We also examine the constrained   NP coupling effects on the dileptonic invariant mass spectra,
the single lepton longitudinal polarization asymmetries
and the forward-backward asymmetries in these decays.

The outline of this paper is follows: In section \ref{THEORETY}, we recapitulate briefly
NP contribution to the LFV exclusive $b \to s \ell_i^- \ell_j^+ $ decays in the scalar LQ model.
In section \ref{Result} we present the numerical analysis for the branching ratios and other physical observed quantities.
Section \ref{Summary} contains the summary and conclusion.

\section{Scalar LQ contributions to the LFV exclusive $b \to s \ell_i^- \ell_j^+$  decays} \label{THEORETY}
The family lepton number ($L_e,L_{\mu},L_{\tau}$) are exactly conserved in the SM,
consequently, the LFV processes are absolutely forbidden in the SM. Nevertheless,
the LFV processes, for an example,
$b \to s \ell^-_i \ell^+_j$,  are allowed in scalar LQ model due to the exchange of LQs.

As discussed in Refs. \cite{Becirevic:2016oho,Becirevic:2017jtw,Mohanta:2013lsa,Sahoo:2015wya,Arnold:2013cva,Sahoo:2015pzk,Mohanta:2016lxs,Li:2016vvp,Li:2016pdv,Wang:2016dne},
out of all possible LQ multiplets,  we will consider the minimal renormalizable scalar LQ models,
containing one single additional representation of $SU(3) \times SU(2) \times U(1)$ and
which do not allow proton decay at the tree level.
The scalar LQ multiplets can have the representation as $X=(3,2,7/6)$ and $X=(3,2,1/6)$
under the gauge group $SU(3) \times SU(2) \times U(1)$.

\begin{figure}[b]
\begin{minipage}{\textwidth}
\centering
\includegraphics[scale=0.5]{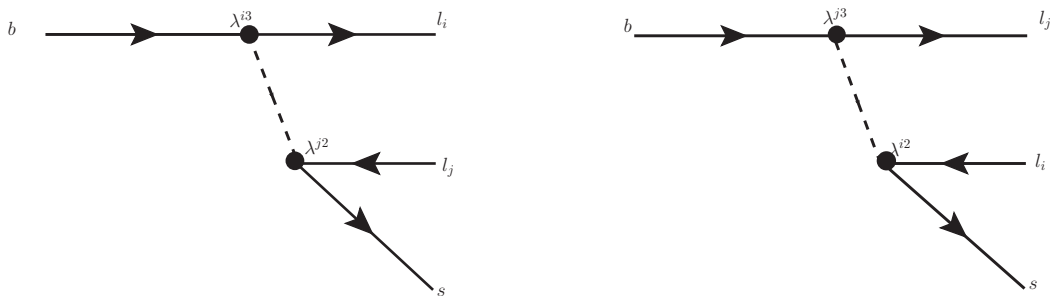}
\caption{Feynman diagrams for the processes  $b \to s \ell_i^- \ell_j^+$
(left) and $b \to s \ell_i^+ \ell_j^-$ (right) due to the scalar LQ exchanges, where $\ell = e, \mu, \tau$.}
 \label{fig:lq}
\end{minipage}
\end{figure}

These scalar LQs can have sizable Yukawa couplings and could potentially
contribute to the quark level transition $b \to q \ell^- \ell^+$.
The tree level Feynman diagram for the LFV process $b \to s \ell^- \ell^+$ are displayed in Fig. \ref{fig:lq}.
Owing to the chirality, diagonality nature and the conservation of both baryon and lepton number,
these LQs processes will provide an effective way looking for their effects in rare $B$ meson decays.
The details of new contributions have been explicitly discussed in Refs. \cite{Mohanta:2013lsa,Sahoo:2015wya,Arnold:2013cva,Mohanta:2016lxs,Sahoo:2015pzk,Becirevic:2016oho,Becirevic:2017jtw}
and here we outline the main points simply.
There are two Lagrangians for process $b \to s \ell^- \ell^+$ due to the couplings of scalar LQs X=(3,2,7/6) and X=(3,2,1/6)
with fermion bilinear, respectively.

\vskip 0.35 cm
\noindent {\bf Model A: X=(3, 2, 7/6)}
\vskip 0.35 cm
\par            
In this model the interaction Lagrangian is given as
\begin{equation}\label{L of 7/6}
{\mathcal L} = - \lambda_{u}^{ij}~ \bar u_R^i X^T \epsilon L_L^j - \lambda_e^{ij}~\bar e_R^i X^\dagger Q_L^j + h.c.,
\end{equation}
where $i,j$ are the generation indices,  $Q_L$ and $L_L$ are the left handed quark and lepton doublets,
$u_R$ and $e_R$ are the right handed up-type quark and charged lepton singlets, and $\epsilon=i \sigma_2$ is a $2\times 2$ matrix.
 More explicitly these multiplets can be found in Refs. \cite{Mohanta:2013lsa,Sahoo:2015wya,Arnold:2013cva,Sahoo:2015pzk,Mohanta:2016lxs,Becirevic:2016oho,Becirevic:2017jtw}.
Using Fierz transformation,
one can obtain the LQ effective Hamiltonian for the process $ b \to s \ell^-_i \ell^+_j$
\begin{equation}\label{Hamiltonian of 7/6}
{\cal H}_{LQ}= \frac{\lambda^{i3} {\lambda^{j2 }}^* }{8 M_Y^2} [ \bar s \gamma^\mu (1-\gamma_5)b]
[\bar \ell_i \gamma_\mu(1+\gamma_5) \ell_j ]=G_{LQ}^{\frac{7}{6}}[ \bar s \gamma^\mu (1-\gamma_5)b]
[\bar \ell_i \gamma_\mu(1+\gamma_5) \ell_j ].
\end{equation}

\vskip 0.4 cm
\noindent {\bf Model B: X=(3, 2, 1/6) }
\vskip 0.35 cm
\par
In this model the interaction Lagrangian is given as
\begin{equation}\label{L of 1/6}
{\cal L} = - \lambda_d^{ij}~ {\bar d}_{\alpha R}^{~i} (V_\alpha e_L^j-Y_\alpha \nu_L^j) +h.c.
\end{equation}
Then the LQ effective Hamiltonian are gotten as
\begin{equation}\label{Hamiltonian of 1/6}
{\cal H}_{LQ}= \frac{\lambda^{2i} {\lambda^{3j }}^* }{8 M_Y^2} [ \bar s \gamma^\mu (1+\gamma_5)b]
[\bar \ell_i \gamma_\mu(1-\gamma_5) \ell_j ]=G_{LQ}^{\frac{1}{6}}[ \bar s \gamma^\mu (1+\gamma_5)b]
[\bar \ell_i \gamma_\mu(1-\gamma_5) \ell_j ].
\end{equation}
From Eq. (\ref{Hamiltonian of 7/6}) and Eq. (\ref{Hamiltonian of 1/6}),
one can see that the difference between  two  current matrix elements is that replace $1\pm \gamma_5$ with $ 1\mp\gamma_5$.
Nevertheless, the different effects only obviously appeared on the single lepton longitudinal polarization asymmetries.

\subsection{The $B_s\to \ell^-_i\ell^+_j$ decays}\label{puredecay}
~~~~Using the above information,
 we can get total decay branching ratios
\begin{eqnarray}\label{Brpure}
\mathcal{B}(B_s \to \ell_i^-  \ell_j^+)=\frac{\tau_{B_s}\sqrt{\lambda(m^2_{B_s},m^2_{\ell_i},m^2_{\ell_j})}}{4\pi m^3_{B_s}}
\bigg\{\big|G_{LQ}f_{Bs}\big|^2 \left[m^2_{B_s}\big(m^2_{\ell_i}+m^2_{\ell_j}\big)-\big(m^2_{\ell_i}-m^2_{\ell_j}\big)^2\right]\bigg\}
\end{eqnarray}
where $G_{LQ}$ is $G_{LQ}^{\frac{1}{6}}$ or $G_{LQ}^{\frac{7}{6}}$,  $\lambda(a,b,c)\equiv a^2+b^2+c^2-2ab-2ac-2bc$,
and this result is consistent with the expression in Ref. \cite{Sahoo:2015wya}.

\subsection{The $B \to K^{(*)} \ell^-_i\ell^+_j$ decays}\label{Semidecay}
~~~~From Eqs. (\ref{Hamiltonian of 7/6}) and (\ref{Hamiltonian of 1/6}), we can  get
the differential decay distribution for these semileptonic decays with respects to $s$
\small{
\begin{eqnarray}\label{dBrdsB2K}
\frac{d\Gamma(B \to K \ell^-_i\ell^+_j)}{ds}&=& \frac{ u(s)|G_{LQ}|^2}{2^{8} \pi^3 m^3_B s} \Bigg\{
                           \bigg[ \big|H_t(q^2)\big|^2 \Big( \Big|h_{\frac{1}{2},\frac{1}{2}}\Big|^2+\Big|h_{-\frac{1}{2},-\frac{1}{2}}\Big|^2\Big)\nonumber\\
                            &&+
                           \frac{1}{3}\big|H_{0}(q^2)\big|^2\Big(\Big|h_{\frac{1}{2},\frac{1}{2}}\Big|^2+\Big|h_{-\frac{1}{2},\frac{1}{2}}\Big|^2 +\Big|h_{\frac{1}{2},-\frac{1}{2}}\Big|^2+\Big|h_{-\frac{1}{2},-\frac{1}{2}}\Big|^2\Big)
                            \bigg]\Bigg\},\\
%
\frac{d{\Gamma}(B \to K^{*} \ell^-_i\ell^+_j)}{ds}&=& \frac{ u(s)|G_{LQ}|^2}{2^{8} \pi^3 m^3_B s} \Bigg\{
                             \bigg[\big|H_{0t}\big|^2\Big(\Big|h_{\frac{1}{2},\frac{1}{2}}\Big|^2 +\Big|h_{-\frac{1}{2},-\frac{1}{2}}\Big|^2\Big) +\frac{1}{3}\Big( \big|H_{00}\big|^2 + \big|H_{++}\big|^2 +\big|H_{--}\big|^2 \Big)\nonumber \\
                             &&\times\Big(\Big|h_{\frac{1}{2},\frac{1}{2}}\Big|^2+ \Big|h_{-\frac{1}{2},\frac{1}{2}}\Big|^2 + \Big|h_{\frac{1}{2},-\frac{1}{2}}\Big|^2  +\Big|h_{-\frac{1}{2},-\frac{1}{2}}\Big|^2\Big)\bigg] \Bigg\},\label{dBrdsB2Ks}
\end{eqnarray}
}with $u(s)=\sqrt{\lambda(m_B^2,m_{K^(*)}^2,s)\lambda(m_{\ell_i}^2,m_{\ell_j}^2,s)}$.

There are other observables investigated due to scalar LQ exchange.
Using the definition formula of forward-backward asymmetries in Refs. \cite{Bensalem:2002ni},
we get the specific expressions of $\mathcal{A}_{FB}(B \to K^{(*)} \ell^-_i\ell^+_j)(s)$ in the scalar LQ model.
\begin{eqnarray}\label{AFBK}
\mathcal{A}_{FB}(B \to K \ell^-_i\ell^+_j)(s) = -\frac{1}{D_K } \big|H_t(q^2)H_{0}(q^2)\big|\Big( \Big|h_{\frac{1}{2},\frac{1}{2}}\Big|^2 + \Big|h_{-\frac{1}{2},-\frac{1}{2}}\Big|^2 \Big) ,
\end{eqnarray}
with
\begin{eqnarray}\label{DK}
D_K &=&\frac{1}{3}\big|H_{0}(q^2)\big|^2\Big(\Big|h_{\frac{1}{2},\frac{1}{2}}\Big|^2+\Big|h_{-\frac{1}{2},\frac{1}{2}}\Big|^2 +\Big|h_{\frac{1}{2},-\frac{1}{2}}\Big|^2+\Big|h_{-\frac{1}{2},-\frac{1}{2}}\Big|^2\Big)\nonumber\\
&&+ \big|H_t(q^2)\big|^2\Big( \Big|h_{\frac{1}{2},\frac{1}{2}}\Big|^2+\Big|h_{-\frac{1}{2},-\frac{1}{2}}\Big|^2\Big).
\end{eqnarray}
For $\mathcal{A}_{FB}(B \to K^{*} \ell^-_i\ell^+_j)$, they are  same with Eq. (11) in Ref. \cite{Sheng:2018ylg} except that
\begin{eqnarray}\label{DKs}
N_{K^*} &=& -2\big|H_{0t}H_{00}\big|\Big(\Big|h_{\frac{1}{2},\frac{1}{2}}\Big|^2+\Big|h_{-\frac{1}{2},-\frac{1}{2}}\Big|^2\Big)+\frac{1}{2}\Big(\big|H_{++}\big|^2 -\big|H_{--}\big|^2 \Big)\bigg( \Big|h_{\frac{1}{2},-\frac{1}{2}}\Big|^2 - \Big|h_{-\frac{1}{2},\frac{1}{2}}\Big|^2 \Big),\nonumber \\
D_{K^*} &=& \frac{2}{3}\Big( \big|H_{00}\big|^2 + \big|H_{++}\big|^2 +\big|H_{--}\big|^2 \Big)\Big(\Big|h_{\frac{1}{2},\frac{1}{2}}\Big|^2+ \Big|h_{-\frac{1}{2},\frac{1}{2}}\Big|^2 + \Big|h_{\frac{1}{2},-\frac{1}{2}}\Big|^2
               +\Big|h_{-\frac{1}{2},-\frac{1}{2}}\Big|^2\Big)\nonumber \\
               &&+2\big|H_{0t}\big|^2\Big(\Big|h_{\frac{1}{2},\frac{1}{2}}\Big|^2 +\Big|h_{-\frac{1}{2},-\frac{1}{2}}\Big|^2\Big).
\end{eqnarray}
Noted that the NP coupling parameter is counteracted  in final result of $\mathcal{A}_{FB}(B \to K^{(*)} \ell^-_i\ell^+_j)(s)$.

Besides, we also study the single lepton  polarization asymmetries.
For the sake of simplicity only the longitudinal component product of single lepton  polarization asymmetries $P^{\ell^\pm}_i(i=L)$ are studied.
The single lepton longitudinal polarization asymmetries are got from \cite{Choudhury:2003mi}.
The specific expressions of $\mathcal{P}^{\ell^{\pm}}_L(B \to K^{(*)} \ell^-_i\ell^+_j)(s)$ in  the scalar LQ model are similar with
Eq. (14) and (17) in Ref. \cite{Sheng:2018ylg}  except $N_K^{'\ell^-},N_K^{'\ell^+},N^{'\ell^-}_{K^*}$ and $N^{'\ell^+}_{K^*}$, which are given as
\begin{eqnarray}
  N_K^{'\ell^-} &=& \bigg\{\frac{1}{3}\big|H_{0}(q^2)\big|^2\Big[\Big( \Big|h_{\frac{1}{2},\frac{1}{2}}\Big|^2-\Big|h_{-\frac{1}{2},-\frac{1}{2}}\Big|^2\Big) +\Big(\Big|h_{\frac{1}{2},-\frac{1}{2}}\Big|^2-\Big|h_{-\frac{1}{2},\frac{1}{2}}\Big|^2\Big)\Big]\nonumber\\
  &&+ \big|H_t(q^2)\big|^2\Big( \Big|h_{\frac{1}{2},\frac{1}{2}}\Big|^2-\Big|h_{-\frac{1}{2},-\frac{1}{2}}\Big|^2\Big)\bigg\},\label{eq.NKm}\\
  N_K^{'\ell^+} &=& \bigg\{\frac{1}{3}\big|H_{0}(q^2)\big|^2\Big[\Big( \Big|h_{\frac{1}{2},\frac{1}{2}}\Big|^2-\Big|h_{-\frac{1}{2},-\frac{1}{2}}\Big|^2\Big) -\Big(\Big|h_{\frac{1}{2},-\frac{1}{2}}\Big|^2-\Big|h_{-\frac{1}{2},\frac{1}{2}}\Big|^2\Big)\Big]\nonumber\\
  && + \big|H_t(q^2)\big|^2\Big( \Big|h_{\frac{1}{2},\frac{1}{2}}\Big|^2-\Big|h_{-\frac{1}{2},-\frac{1}{2}}\Big|^2\Big)\bigg\},\label{eq.NKp}\\
 N^{'\ell^-}_{K^*} &=&  \bigg\{\frac{2}{3}\Big( \big|H_{00}\big|^2 + \big|H_{++}\big|^2 +\big|H_{--}\big|^2 \Big)\Big[\Big(\Big|h_{\frac{1}{2},\frac{1}{2}}\Big|^2- \Big|h_{-\frac{1}{2},-\frac{1}{2}}\Big|^2\Big) +  \Big(\Big|h_{\frac{1}{2},-\frac{1}{2}}\Big|^2
-\Big|h_{-\frac{1}{2},\frac{1}{2}}\Big|^2\Big)\Big]\nonumber \\
&&+2\big|H_{0t}\big|^2\Big(\Big|h_{\frac{1}{2},\frac{1}{2}}\Big|^2 -\Big|h_{-\frac{1}{2},-\frac{1}{2}}\Big|^2\Big)\bigg\},\label{eq.NKVm}\\
 N^{'\ell^+}_{K^*} &=&  \bigg\{\frac{2}{3}\Big( \big|H_{00}\big|^2 + \big|H_{++}\big|^2 +\big|H_{--}\big|^2 \Big)\Big[\Big(\Big|h_{\frac{1}{2},\frac{1}{2}}\Big|^2- \Big|h_{-\frac{1}{2},-\frac{1}{2}}\Big|^2\Big) - \Big( \Big|h_{\frac{1}{2},-\frac{1}{2}}\Big|^2
-\Big|h_{-\frac{1}{2},\frac{1}{2}}\Big|^2\Big)\Big]\nonumber \\
&&+2\big|H_{0t}\big|^2\Big(\Big|h_{\frac{1}{2},\frac{1}{2}}\Big|^2 -\Big|h_{-\frac{1}{2},-\frac{1}{2}}\Big|^2\Big)\bigg\}.\label{eq.NKVp}
\end{eqnarray}

\section{Numerical results and analysis}\label{Result}

In this section, we give our numerical results  of the scalar LQ contributions in  LFV $B_s\to \ell^-_i \ell^+_j$ and $B \to K^{(*)} \ell^-_i \ell^+_j$ decays. The relevant input parameters and form factors are mainly from Refs. \cite{PDG2016,Ball:2004rg,Wu:2006rd}.
In our analysis,  we will use the latest LFV experimental upper limits of $\mathcal{B}(B_s\to \ell^-_i \ell^+_j)$
and $\mathcal{B}(B \to K^{(*)} \ell^-_i \ell^+_j)$  listed in the second column of Tab. \ref{tab:BrLQPre}, which is different from  Refs. \cite{Sahoo:2015wya,Mohanta:2013lsa,Sahoo:2015pzk} by using the experimental $\mathcal{B}(B_s \to \mu^- \mu^+)$, to constrain the relevant LQ parameters.
\begin{table}[th]            \renewcommand{\arraystretch}{0.915}
\caption{\footnotesize{The experimental measurements  \cite{PDG2016} and our scalar LQ predictions for exclusive $\bar{b}\to \bar{s} \ell_i^-\ell_j^+$ decays.}}\label{tab:BrLQPre}
\begin{center}{\small
\begin{tabular}{lcc}
\hline\hline
 ~~~~~~Observables & Exp. Limits  at $90\%$ C.L.  &~~Our Scalar LQ Predictions 
 \\\hline
$\mathcal{B}(B_s\to e^{-}\mu^{+})(\times10^{-8})$&$\leqslant1.1$&$\leqslant0.04$ \\
$\mathcal{B}(B^0_d\to  K^0 e^-\mu^+)(\times10^{-7})$ & $\leqslant2.7$ &$\leqslant1.20$\\
$\mathcal{B}(B^+_u\to  K^+ e^-\mu^+)(\times10^{-7})$ &$\leqslant1.3$ &{\color{blue}$\leqslant1.30$} \\
$\mathcal{B}(B^0_d\to  K^{*0} e^-\mu^+)(\times10^{-7})$&  $\leqslant3.4$ &{\color{blue}$\leqslant3.40$} \\
$\mathcal{B}(B^+_u\to  K^{*+} e^-\mu^+)(\times10^{-7})$ &  $\leqslant9.9$ &$\leqslant3.69$ \\
\hline
$\mathcal{B}(B_s\to \mu^{-}e^{+})(\times10^{-8})$&$\leqslant1.1$&$\leqslant0.03$ \\
$\mathcal{B}(B^0_d\to  K^0 \mu^-e^+)(\times10^{-7})$&$\leqslant2.7$ &$\leqslant0.85$ \\
$\mathcal{B}(B^+_u\to  K^+ \mu^-e^+)(\times10^{-8})$&$\leqslant9.1$&{\color{blue}$\leqslant9.10$} \\
$\mathcal{B}(B^0_d\to  K^{*0} \mu^-e^+)(\times10^{-7})$&$\leqslant5.3$ &$\leqslant3.79$ \\
$\mathcal{B}(B^+_u\to  K^{*+} \mu^-e^+)(\times10^{-6})$& $\leqslant1.3$&$\leqslant0.41$ \\
\hline
$\mathcal{B}(B_s\to e^{-}\tau^{+})(\times10^{-5})$  &...&$\leqslant1.65$ \\
$\mathcal{B}(B^0_d\to  K^0 e^-\tau^+)(\times10^{-5})$ &...&$\leqslant1.39$ \\
$\mathcal{B}(B^+_u\to  K^+ e^-\tau^{+})(\times10^{-5})$ & $\leqslant1.5$ &{\color{blue}$\leqslant1.50$} \\
$\mathcal{B}(B^0_d\to  K^{*0} e^-\tau^+)(\times10^{-5})$ & ...  &$\leqslant4.11$ \\
$\mathcal{B}(B^+_u\to  K^{*+} e^-\tau^{+})(\times10^{-5})$ & ...  &$\leqslant4.57$ \\
\hline
$\mathcal{B}(B_s\to \tau^{-}e^{+})(\times10^{-5})$&...&$\leqslant5.23$ \\
$\mathcal{B}(B^0_d\to  K^0 \tau^-e^+)(\times10^{-5})$    &   ...   &$\leqslant3.99$ \\
$\mathcal{B}(B^+_u\to  K^+ \tau^-e^{+})(\times10^{-5})$  &  $\leqslant4.3$ &{\color{blue}$\leqslant4.30$} \\
$\mathcal{B}(B^0_d\to  K^{*0} \tau^-e^+)(\times10^{-5})$      &   ...   &$\leqslant12.99$  \\
$\mathcal{B}(B^+_u\to  K^{*+} \tau^-e^{+})(\times10^{-5})$    &   ...   &$\leqslant12.09$  \\
\hline
$\mathcal{B}(B_s\to \mu^{-}\tau^{+})(\times10^{-5})$&...&$\leqslant3.25$   \\
$\mathcal{B}(B^0_d\to  K^0 \mu^-\tau^+)(\times10^{-5})$  &   ...   &$\leqslant2.60$  \\
$\mathcal{B}(B^+_u\to  K^+ \mu^-\tau^{+})(\times10^{-5})$&  $\leqslant2.8$ &{\color{blue}$\leqslant2.80$ } \\
$\mathcal{B}(B^0_d\to  K^{*0} \mu^-\tau^+)(\times10^{-5})$    &   ...   &$\leqslant8.21$ \\
$\mathcal{B}(B^+_u\to  K^{*+} \mu^-\tau^{+})(\times10^{-5})$  &   ...   &$\leqslant8.31$  \\
\hline
$\mathcal{B}(B_s\to \tau^{-}\mu^{+})(\times10^{-5})$&...&$\leqslant5.47$ \\
$\mathcal{B}(B^0_d\to  K^0 \tau^-\mu^+)(\times10^{-5})$&...&$\leqslant4.18$  \\
$\mathcal{B}(B^+_u\to  K^+ \tau^-\mu^{+})(\times10^{-5})$&$\leqslant4.5$ &{\color{blue}$\leqslant4.50$}  \\
$\mathcal{B}(B^0_d\to  K^{*0} \tau^-\mu^+)(\times10^{-5})$&...&$\leqslant12.50$  \\
$\mathcal{B}(B^+_u\to  K^{*+} \tau^-\mu^{+})(\times10^{-5})$& ...&$\leqslant14.60$  \\
\hline
\end{tabular}}
\end{center}
\end{table}

\begin{table}[htb]   
\caption{\footnotesize{Upper limits of the relevant LQ coupling parameters constrained from $B_s\to \ell^-_i \ell^+_j$ and
$B \to K^{(*)} \ell^-_i\ell^+_j$ LFV decays (in units of {\scriptsize{$\rm GeV^{-2}$}}),
and previous strongest bounds \cite{Davidson:1993qk,Benbrik:2010cf} are listed for comparison.}}\label{tab:NP coupling parameters}
\begin{center}
\begin{tabular}{|l|c|l|c|}
\hline
 LFV couplings & ~Relevant processes& Our bounds&~Previous  bounds 
\\ \hline
Model A~~$\frac{|\lambda^{13}{\lambda}^{22*}|}{M_s^2}$& & &\small{$3\times10^{-7}$}\small{$[B\to K \bar{e}\mu]$}\\[0.3ex]
&\raisebox{2.5ex}{$b \to s e^-\mu^+$}   & \raisebox{2.5ex}{$\leqslant6.61\times10^{-9}$} &  \\[-4.5ex]
Model B~~$\frac{|\lambda^{21}{\lambda}^{32*}|}{M_s^2}$&  &  &\small{${2\times10^{-9}}[K\to \pi\nu \bar{\nu}]$}\\
\hline
Model A~~$\frac{|\lambda^{23}{\lambda}^{12*}|}{M_s^2}$& & &\small{$3\times10^{-7}[B\to K \bar{e}\mu]$}\\[0.3ex]
&\raisebox{2.5ex}{$b \to s \mu^-e^+$}   & \raisebox{2.5ex}{$\leqslant5.57\times10^{-9}$} &          \\[-4.5ex]
Model B~~$\frac{|\lambda^{22}{\lambda}^{31*}|}{M_s^2}$&  &  &\small{${2\times10^{-9}}[K\to \pi\nu \bar{\nu}]$} \\
\hline
Model A~~$\frac{|\lambda^{13}{\lambda}^{32*}|}{M_s^2}$& &   &
\small{ ${4\times10^{-6}}[B\to X \ell \nu]$} \\[0.3ex]
   &\raisebox{2.5ex}{$b \to s  e^-\tau^+$}  &\raisebox{2.5ex}{$\leqslant 9.17\times10^{-8}$}&  \\[-4.5ex]
Model B~~$\frac{|\lambda^{21}{\lambda}^{33*}|}{M_s^2}$&  &  &
\small{${2\times10^{-7}}[B\to X \tau \bar{\mu}]$} \\
\hline
Model A~~$\frac{|\lambda^{33}{\lambda}^{12*}|}{M_s^2}$& &   &\small{${4\times10^{-6}}[B\to X \ell \nu]$} \\[0.3ex]
&\raisebox{2.5ex}{$b \to s \tau^- e^+$} & \raisebox{2.5ex}{$\leqslant1.59\times10^{-7}$} &  \\[-4.5ex]
Model B~~$\frac{|\lambda^{23}{\lambda}^{31*}|}{M_s^2}$& &     &\small{$2\times10^{-7}[B\to X \tau \bar{\mu}]$}\\
\hline
Model A~~$\frac{|\lambda^{23}{\lambda}^{32*}|}{M_s^2}$&&& \\[0.3ex]
&\raisebox{2.5ex}{$b \to s \mu^-\tau^+$} &  \raisebox{2.5ex}{$\leqslant1.28\times10^{-7}$}
&\raisebox{2.5ex}{\small{$2\times10^{-7}[B\to X \tau \bar{\mu}]$}} \\[-4.5ex]
Model B~~$\frac{|\lambda^{22}{\lambda}^{33*}|}{M_s^2}$& & & \\
\hline
Model A~~$\frac{|\lambda^{33}{\lambda}^{22*}|}{M_s^2}$ &   &  &\\[0.3ex]
&\raisebox{2.5ex}{$b \to s  \tau^-\mu^+$}   & \raisebox{2.5ex}{$\leqslant1.63\times10^{-7}$}
&\raisebox{2.5ex}{\small{$2\times10^{-7}[B\to X \tau \bar{\mu}]$}} \\[-4.5ex]
Model B~~$\frac{|\lambda^{23}{\lambda}^{32*}|}{M_s^2}$&  &  & \\
\hline
\end{tabular}
\end{center}
\end{table}

First, using the recent 90\% C.L.  experimental measurements of  $\mathcal{B}(B_s\to \ell^-_i \ell^+_j)$ and
$\mathcal{B}(B \to K^{(*)} \ell^-_i\ell^+_j)$ listed in the second column of Tab. \ref{tab:BrLQPre}, we obtain
the upper limits  of  the relevant LQ couplings, which are given in Tab. \ref{tab:NP coupling parameters}. The previous strongest  upper limits are also listed in the last column of Tab. \ref{tab:NP coupling parameters} for convenient comparison.
From Tab. \ref{tab:NP coupling parameters}, one can see  that the  moduli of the constrained LQ coupling products appeared in
these six processes are strongly constrained by present experimental upper limits of relevant  LFV decays.
For  $\frac{|\lambda^{21}{\lambda}^{32*}|}{M_s^2}$ and $\frac{|\lambda^{22}{\lambda}^{31*}|}{M_s^2}$  couplings in Model B, which are related to $b \to s e^-\mu^+$ and  $b \to s \mu^-e^+$, respectively,  our bounds  are slight weaker than previous ones from $K\to \pi\nu\bar{\nu}$. For other LQ couplings,  our bounds are stronger than previous ones, and our some bounds are stronger about one or two orders of magnitude than previous ones in Refs. \cite{Davidson:1993qk,Benbrik:2010cf}.

Next, using the constrained spaces listed in Tab. \ref{tab:NP coupling parameters}, we investigate the scalar LQ coupling effects on the branching ratios, the dileptonic invariant mass spectra, the differential forward-backward asymmetries
and the single lepton longitudinal polarization asymmetries
of relevant $B$ meson LFV decays.
 Noted that, we find that  all predictions except the single lepton longitudinal polarization asymmetries $(P_L^{\ell^{\pm}})$ are similar  in both model A and model B, which are consistent with the results in Refs. \cite{Mohanta:2013lsa,Sahoo:2015wya,Arnold:2013cva,Sahoo:2015pzk,Becirevic:2017jtw}, so the same results between model A and model B will be labeled  scalar LQ predictions in following text.

Our numerical results of the branching ratios considering the LFV scalar LQ contributions  are listed in the last column of Tabs. \ref{tab:BrLQPre}. Comparing with  the corresponding  experimental bounds,
one can find that experimental bounds of $\mathcal{B}(B^0_d \to K^{*0} e^- \mu^+)$ and all $\mathcal{B}(B^+_u \to K^+ \ell^-_i \ell^+_j)$ give effective constraints on the relevant LQ couplings.
In addition, we also explore the sensitivities of  the  branching ratios to the scalar  LQ couplings as well as the scalar LQ coupling effects on  the dileptonic invariant mass spectra and the differential forward-backward asymmetries.
Since their predictions  are similar with our previous results in Ref. \cite{Sheng:2018ylg}, which are considered the R-parity violating couplings, we will not show them again in this paper.

\begin{figure}[b]
  \centering
\includegraphics[scale=0.72]{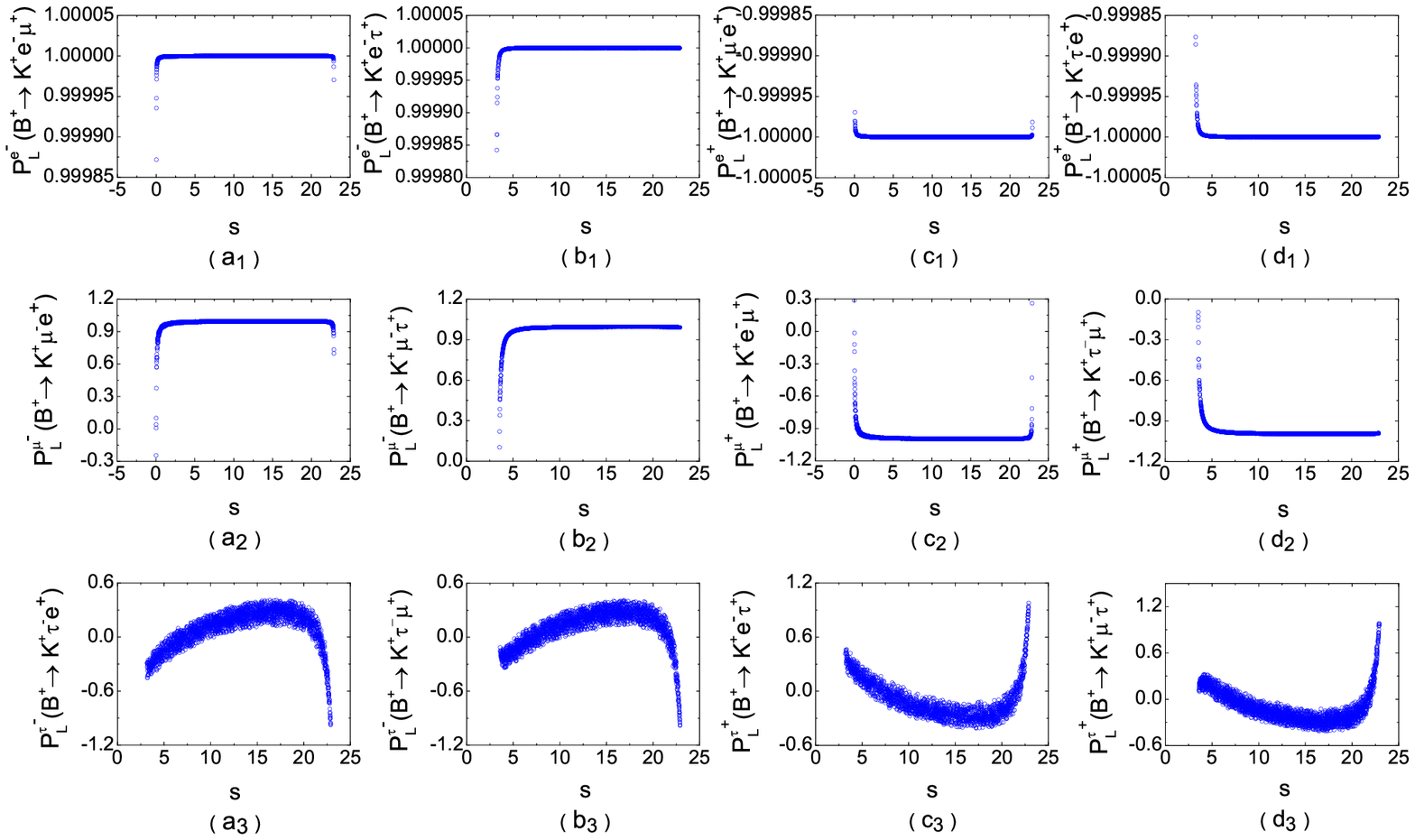}
 \caption{\footnotesize{The  predictions of  $P_L^{\ell^{\pm}}(B^+ \to K^{+} \ell^-_i \ell^+_j)$ decays in model A X=(3,2,7/6).}}
 \label{fig:7PK}
  \centering
  \vspace{0.3cm}
\includegraphics[scale=0.72]{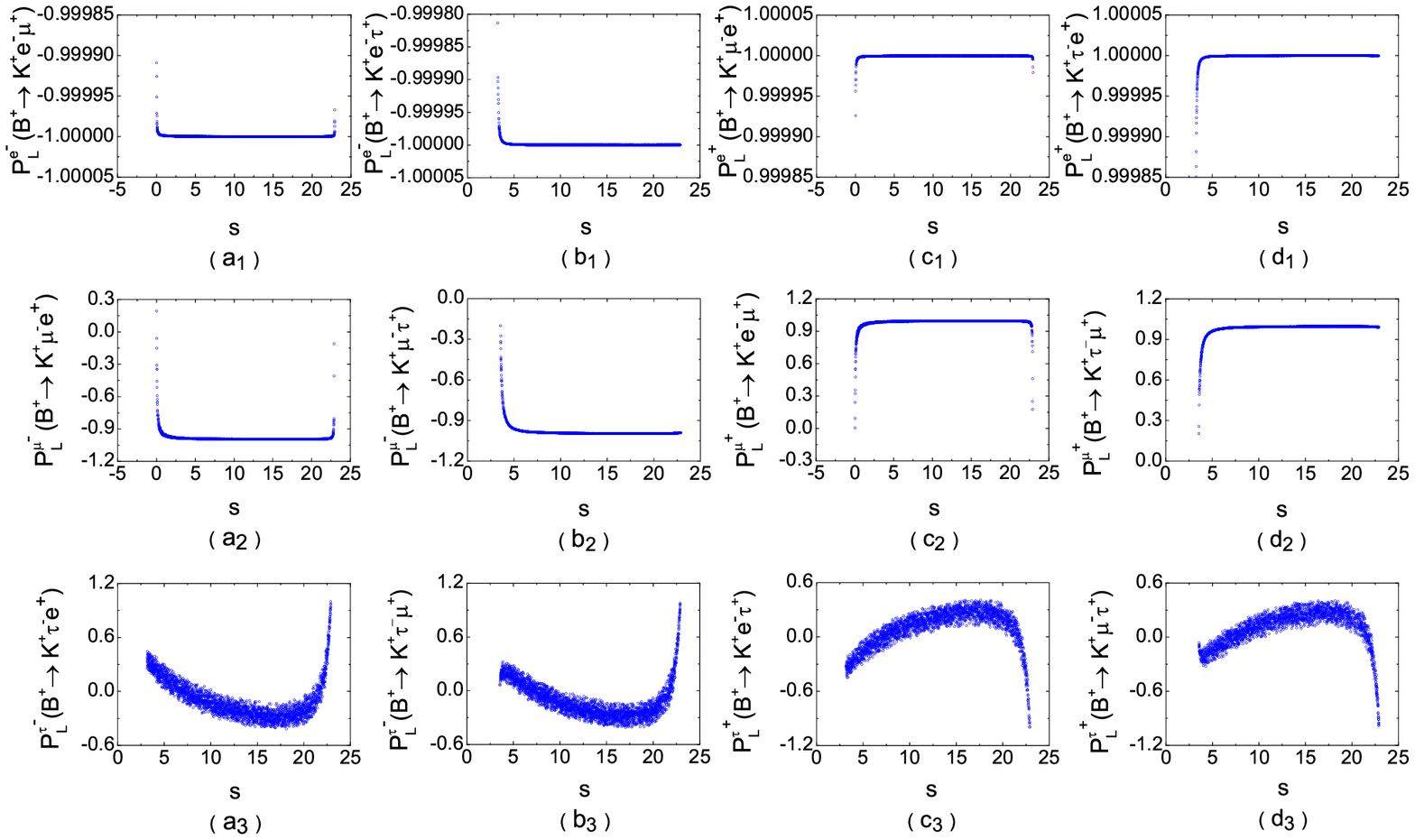}
 \caption{\footnotesize{The  predictions of $P_L^{\ell^{\pm}}(B^+ \to K^{+} \ell^-_i \ell^+_j)$ decays in model B X=(3,2,1/6).}}
 \label{fig:1PK}
\end{figure}

\begin{figure}[htbp]
\vspace{0.5cm}
  \centering
\includegraphics[scale=0.75]{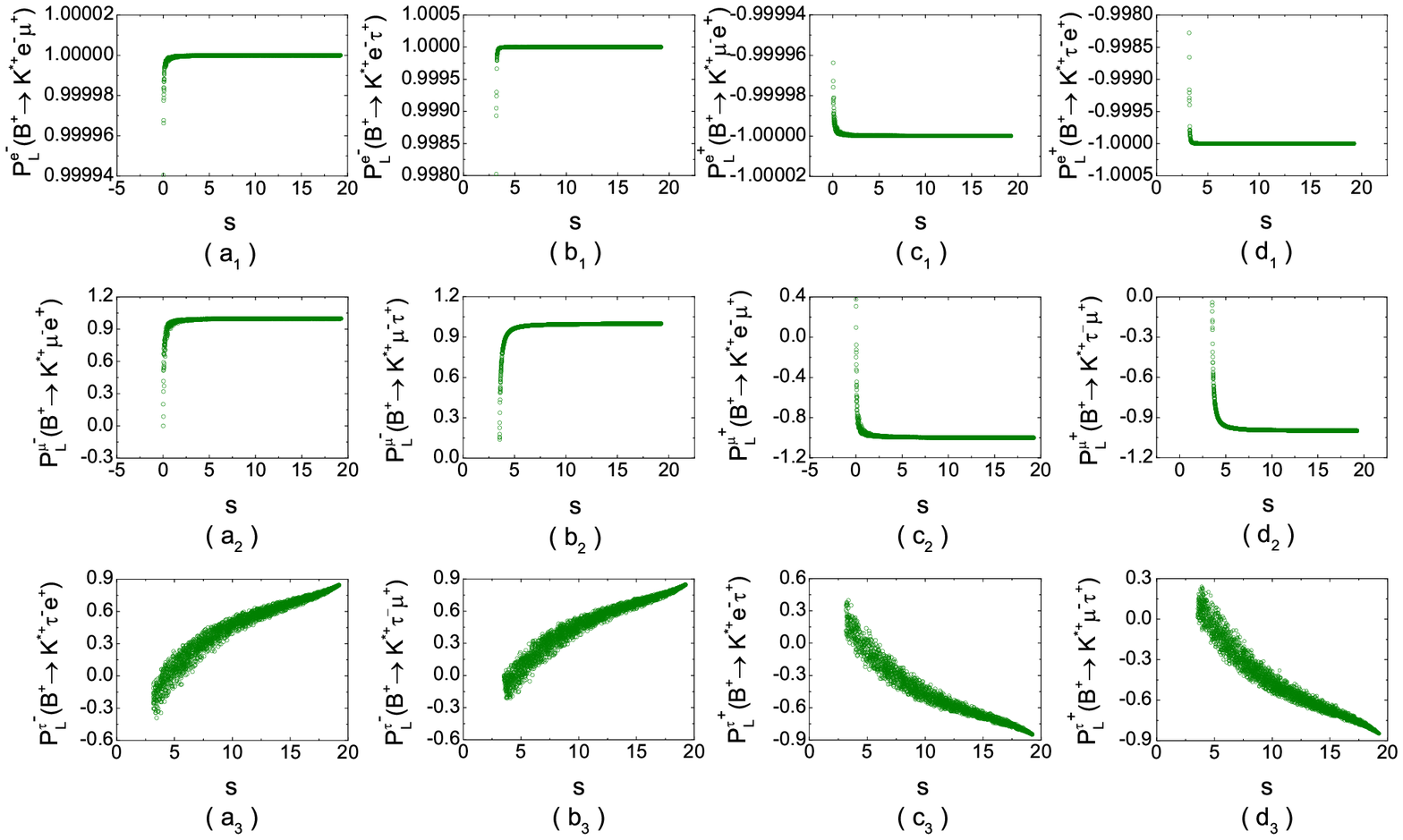}
 \caption{\footnotesize{The predictions of  $P_L^{\ell^{\pm}}(B^+ \to K^{*+} \ell^-_i \ell^+_j)$ decays in model A X=(3,2,7/6).}}
 \label{fig:7PKs}
\vspace{0.5cm}
  \centering
\includegraphics[scale=0.75]{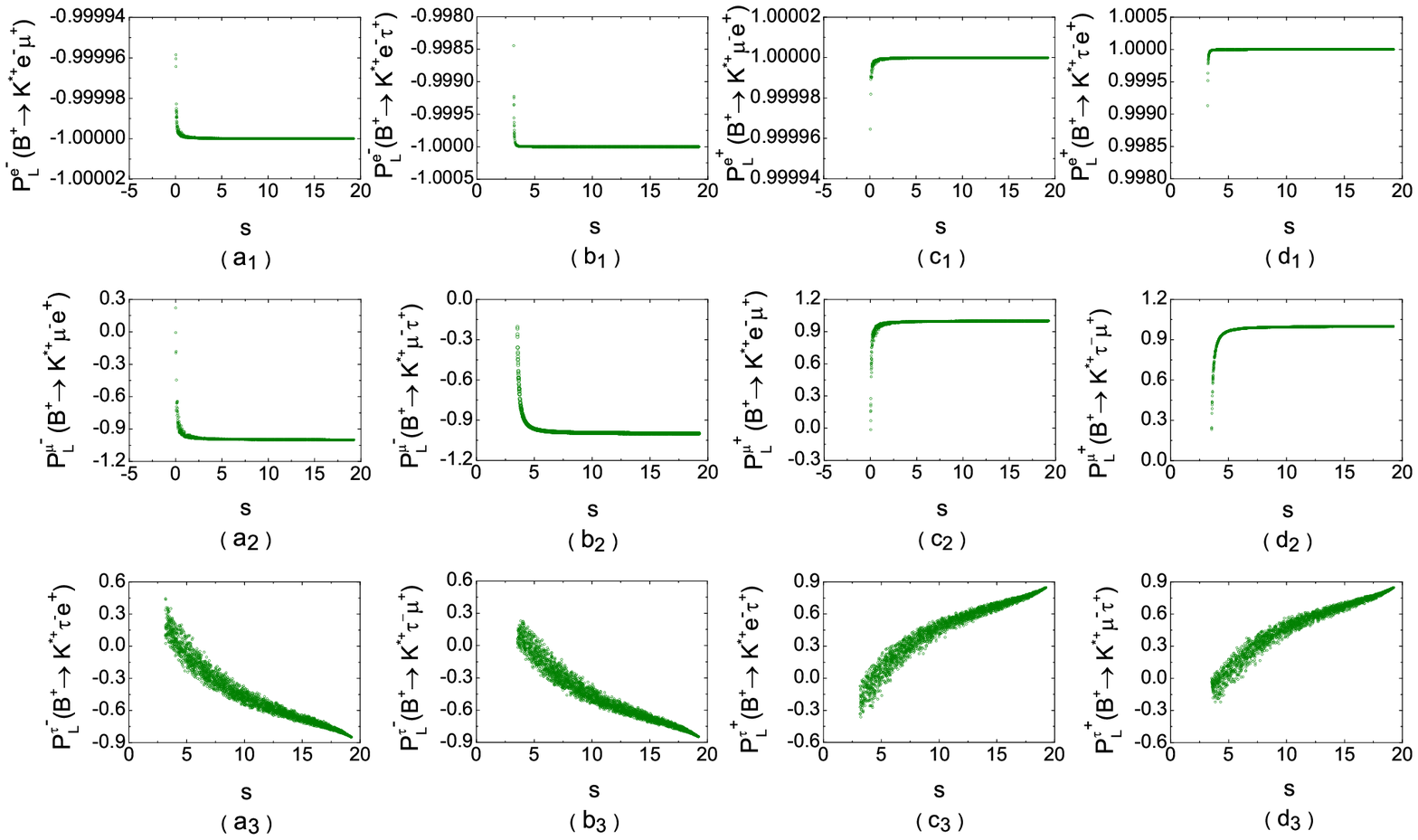}
 \caption{\footnotesize{The predictions of $P_L^{\ell^{\pm}}(B^+ \to K^{*+} \ell^-_i \ell^+_j)$  in model B X=(3,2,1/6).}}
 \label{fig:1PKs}
\end{figure}

Now we give the predictions of  the single lepton longitudinal polarization asymmetries ($P_L^{\ell^{\pm}}$).
Figs. \ref{fig:7PK}$-$\ref{fig:1PKs} show our  predictions of  $P_L^{\ell^{\pm}}$ in the model A and model B,
and we can see that the effects of the scalar LQ exchange couplings are quite different between model A and model B, and the predictions of $P_L^{\ell^{\pm}}$ in model A have exactly the reverse effects with ones in model B. From Eq. (\ref{h17}), one can see that the difference is due to  the lepton section. These differences could be used to distinguish between model A and model B.

Noted that, although the branching ratios and the dileptonic invariant mass spectra of the LFV semileptonic $B \to K^{(*)} \ell^-_i \ell^+_j$ decays have been  investigated in the scalar LQ model in many Refs. \cite{Mohanta:2013lsa,Sahoo:2015wya,Arnold:2013cva,Sahoo:2015pzk,Becirevic:2017jtw}, the differential forward-backward asymmetries
and the single lepton longitudinal polarization asymmetries of relevant LFV $B$ decays in the scalar LQ model are studied for the first time.

\section{Summary and conclusions}\label{Summary}
In this paper, inspired by the recent anomaly measurements of the LFV decays $h\to\mu\tau$, $\mu \to e \gamma$ and $Z\to e\mu(e\tau,\mu\tau)$
and the lepton flavor non-universality in  $b \to s   \ell^-\ell^+$ decays,  we have investigated the scalar LQ exchange effects
in   $B_s\to \ell^-_i \ell^+_j$ and  $B\to K^{(*)} \ell^-_i \ell^+_j$ LFV processes.
Using the latest experimental measurements  of $\mathcal{B}(B_s \to \ell^-_i \ell^+_j)$ and  $\mathcal{B}(B \to K^{(*)} \ell_i^-\ell_j^+)$ as well as considering the $90\%$ C.L. theoretical uncertainties of the input parameters, we have obtained very strong bounds on the moduli of the scalar LQ coupling parameters involving in  $b \to s  \ell_i^-\ell_j^+$ transitions, and
our most bounds are stronger than previous ones.

Using our constrained  spaces of the scalar LQ couplings, we have predicted the possible LQ coupling effects in relevant LFV $B$ decays. We have found that all branching ratios are great increasing with the moduli of the LFV LQ coupling products, and the LQ coupling predictions of the dileptonic invariant mass spectra as well as the differential forward-backward asymmetries are similar with our previous R-parity violating results  \cite{Sheng:2018ylg}. In addition, the LQ coupling effects on relevant single lepton longitudinal polarization asymmetries are quite different between model A and model B.
 Noted that the differential forward-backward asymmetries
and the single lepton longitudinal polarization asymmetries of relevant LFV $B$ decays in the scalar LQ model have been studied for the first time.

With the rapid development of  LHCb and forthcoming Belle-II
experiments, the results  in this work could be a possible way to  probe the NP effects, and will correlate strongly with study for the LQ signals in the future experiments.

\section*{Acknowledgments}
The work was supported by the National Natural Science Foundation of
China(Contract Nos.  11675137,  11225523, 11775092 and 11047145), Nanhu Scholars Program and the High Performance Computing Lab of Xinyang Normal University.

\begin{appendix}\label{Appendix}

\section*{\centering Appendix}
From Eq. (\ref{Hamiltonian of 7/6}) and Eq. (\ref{Hamiltonian of 1/6}) in Sec. \ref{THEORETY}, we get the following expressions about the square matrix element for $B\to K^{(*)} \ell^-_i \ell^+_j$ processes
\begin{eqnarray}\label{eq:M btos}
\big|{\mathcal{ M}}(B \to K^{(*)} \ell^-_i \ell^+_j)\big|^2=
\big|G_{LQ}\big|^2 \big|\langle K^{(*)}|\bar{s}\gamma^{\mu}(1+\gamma_5){b}|{B} \rangle
\bar{\ell_i}\gamma_{\mu}(1-\gamma_5)\ell_j\big|^2 \equiv \big|G_{LQ}\big|^2 L_{\mu\nu}H^{\mu\nu},
\end{eqnarray}
with $G_{LQ}$ is $G_{LQ}^{7/6}$ and $G_{LQ}^{1/6}$ for model A and model B, respectively.

\section{Formulae of the $ B \to K \ell_i^- \ell_j^+$ decays  }  \label{subsec:BK}
~~~~Because the axial-vector current matrix elements
$\langle K(p_K)|\bar s \gamma^{\mu} \gamma_5  b| B(p_B)\rangle =0$, the hadronic  helicity amplitudes are same with each other in model A and model B.
The non-vanishing helicity amplitudes $H_0(q^2)=\frac{2 m_B |\vec{\textbf{p}}|}{\sqrt{q^2}}f_+(q^2)$ and
$H_{t}(q^2) =\frac{m_B^2 - m_K^2}{\sqrt{q^2}} f_0(q^2)$ with $|\vec{\textbf{p}}|=\frac{\sqrt{\lambda(m^2_B,m_K^2, s)}}{2 m_B}$.
The form factors $f_+(q^2)$ and $f_0(q^2)$ in $H_{0(t)}(q^2)$ are taken  from Refs. \cite{Ball:2004rg,Wu:2006rd,Gratrex:2015hna}.

Using the similar method in Refs. \cite{Korner:1987kd,Kadeer:2005aq},
we obtain the  lepton helicity amplitudes in the LQ model
\begin{eqnarray}\label{hll}
&&h^7_{\mp\frac{1}{2},\mp\frac{1}{2}}=\bar{u}_{\ell_i}(\pm{\frac{1}{2}})\gamma^\mu(1+\gamma_5)v_{\ell_j}(\pm{\frac{1}{2}})
       \left\{{\epsilon_\mu (\pm1) \atop \epsilon_\mu (t),\epsilon_\mu (0)} \right\}\ ,\nonumber  \\
&&h^1_{\mp\frac{1}{2},\mp\frac{1}{2}}=\bar{u}_{\ell_i}(\pm{\frac{1}{2}})\gamma^\mu(1-\gamma_5)v_{\ell_j}(\pm{\frac{1}{2}})
       \left\{{\epsilon_\mu (\pm1) \atop \epsilon_\mu (t),\epsilon_\mu (0)} \right\}\ ,
\end{eqnarray}
where the  superscripts of $h^{7,1}_{\lambda_{\ell_i},\lambda_{\ell_j}}$ represent model A $X=(3,2,7/6)$ and model B $X=(3,2,1/6)$. In model B,  the lepton helicity amplitudes
are same with $h^1$ listed in Eq. (28) due to squark exchange in RPV SUSY model \cite{Sheng:2018ylg}.
In model A, the lepton helicity amplitudes are
\begin{eqnarray}\label{h17}
 \Big|h^7_{~~\frac{1}{2},~~\frac{1}{2}}\Big|^2 &=& \frac{1}{s}\Big[s^2-\Big(m^2_{\ell_j}-m^2_{\ell_i}-\sqrt{\lambda_L}\Big)^2\Big], \nonumber  \\
 \Big|h^7_{~~\frac{1}{2},-\frac{1}{2}}\Big|^2 &=& 4\Big(s-m^2_{\ell_i}-m^2_{\ell_j}+\sqrt{\lambda_L}\Big),  \nonumber \\
 \Big|h^7_{-\frac{1}{2},~~\frac{1}{2}}\Big|^2 &=&  4\Big(s-m^2_{\ell_i}-m^2_{\ell_j}-\sqrt{\lambda_L}\Big), \nonumber  \\
 \Big|h^7_{-\frac{1}{2},-\frac{1}{2}}\Big|^2 &=&\frac{1}{s}\Big[s^2-\big(m^2_{\ell_i}-m^2_{\ell_j}-\sqrt{\lambda_L}\big)^2\Big],
 \end{eqnarray}
with $\lambda_L\equiv\lambda(m^2_{\ell_i},m^2_{\ell_j},s)$.

For $ B \to K  \ell_i^- \ell_j^+ $ processes,
the double differential decay rates with $\lambda_{\ell_{i,j}}$ can be represented as
\begin{eqnarray} \label{eq:double-B2K}
\frac{d^2 \Gamma^K[\lambda_{\ell_i} = \frac{1}{2},\lambda_{\ell_j} = \frac{1}{2}]}{ds d \cos\theta}    &=&  \frac{u(s)\big|G_{LQ}\big|^2}{2^{9} \pi^3 m^3_B s}
                \bigg\{\Big|h_{\frac{1}{2},\frac{1}{2}}\Big|^2  \big|H_t(q^2)- H_{0}(q^2)\cos\theta \big|^2 \bigg\}, \nonumber \\
\frac{d^2 \Gamma^K[\lambda_{\ell_i} =-\frac{1}{2},\lambda_{\ell_j} = \frac{1}{2}]}{ds d \cos\theta} &=&  \frac{u(s)\big|G_{LQ}\big|^2}{2^{9} \pi^3 m^3_B s}
                \bigg\{ \frac{1}{2}\Big|h_{-\frac{1}{2},\frac{1}{2}}\Big|^2  \big|H_{0}(q^2)\big|^2 \sin^2\theta\bigg\},\nonumber \\
\frac{d^2 \Gamma^K[\lambda_{\ell_i} =\frac{1}{2},\lambda_{\ell_j} =-\frac{1}{2}]}{ds d \cos\theta} &=&  \frac{u(s)\big|G_{LQ}\big|^2}{2^{9} \pi^3 m^3_B s}
                \bigg\{  \frac{1}{2} \Big|h_{\frac{1}{2},-\frac{1}{2}}\Big|^2  \big|H_{0}(q^2)\big|^2 \sin^2\theta \bigg\},\nonumber \\
\frac{d^2 \Gamma^K[\lambda_{\ell_i} =-\frac{1}{2},\lambda_{\ell_j} =-\frac{1}{2}]}{ds d \cos\theta}    &=&  \frac{u(s)\big|G_{LQ}\big|^2}{2^{9} \pi^3 m^3_B s} \bigg\{
                \Big|h_{-\frac{1}{2},-\frac{1}{2}}\Big|^2  \big|H_t(q^2)-  H_{0}(q^2)\cos\theta \big|^2 \bigg\},
\end{eqnarray}
where $h_{m,n}$ is $h_{m,n}^7$ and $h_{m,n}^1$ ($G_{LQ}$ is $G_{LQ}^{\frac{1}{6}}$ and $G_{LQ}^{\frac{7}{6}}$)   for model A and model B, respectively.

\section{Formulae of the $B \to K^* \ell_i^- \ell_j^+$ decays}  \label{subsec:BKstar}
Unlike the process $ B\to K$, the hadronic  helicity amplitudes of $B\to K^*$ are different from each other between model A  and model B,
since the $ B \to K^*$ matrix element for the axial-vector current don't vanishes, i.e.,
$\langle K^*(p_{K^*})|\bar s \gamma^{\mu} \gamma_5  b| B(p_B)\rangle \neq0$.
In model B, the hadronic helicity amplitudes
are same with $H^1(q^2)$ in Ref. \cite{Sheng:2018ylg}.
In model A, the helicity amplitudes can be written as
\begin{eqnarray}\label{HV}
H^7_{\pm \pm}(q^2) &=&-(m_B + m_{K^*}) A_1(q^2) \mp \frac{2 m_B}{m_B + m_{K^*}} |\vec{\textbf{p}}|\, V(q^2), \nonumber \\[0.2cm]
H^7_{00}(q^2) &=&-\frac{1}{2 m_{K^*} \sqrt{q^2}} \left[(m_B^2 - m_{K^*}^2 -q^2) (m_B + m_{K^*}) A_1(q^2) - \frac{4 m_B^2 |\vec{\textbf{p}}|^2}{m_B + m_{K^*}} A_2(q^2) \right],
\nonumber \\[0.2cm]
H^7_{0t}(q^2) &=&-\frac{2 m_B |\vec{\textbf{p}}|}{\sqrt{q^2}} A_0(q^2),
\end{eqnarray}
with $|\vec{\textbf{p}}|=\frac{\sqrt{\lambda(m^2_B,m_{K^{*}}^2,s)}}{2 m_B}$.
Noted that the  hadronic helicity amplitudes in model A and model B are opposite.

For $ B \to K^*  \ell_i^- \ell_j^+ $ processes,
the double differential decay rates with $\lambda_{\ell_{i,j}}$ can be represented as
\begin{eqnarray} \label{eq:double-B2Ks}
\frac{d^2 \Gamma^{K^*}[\lambda_{\ell_i} = \frac{1}{2} ,\lambda_{\ell_j} = \frac{1}{2}]}{ds d \cos\theta} &=&  \frac{u(s)\big|G_{LQ}\big|^2}{2^{9} \pi^3 m^3_B s} \bigg\{
                  \Big|h_{\frac{1}{2},\frac{1}{2}}\Big|^2 \bigg[\big|H_{0t}-  H_{00}\cos\theta\big|^2  \nonumber \\
                  && +\frac{1}{2}|H_{++}|^2\sin^2\theta + \frac{1}{2}\big|H_{--}\big|^2\sin^2\theta \bigg] \bigg\},\nonumber \\
\frac{d^2 \Gamma^{K^*}[\lambda_{\ell_i} =-\frac{1}{2},\lambda_{\ell_j} = \frac{1}{2}]}{ds d \cos\theta} &=&  \frac{u(s)\big|G_{LQ}\big|^2}{2^{9} \pi^3 m^3_B s} \bigg\{
                 \Big|h_{-\frac{1}{2},\frac{1}{2}}\Big|^2 \bigg[\frac{1}{2} \big|H_{00}\big|^2\sin^2\theta  \nonumber \\
                  &&+ \frac{1}{4}\big|H_{++}\big|^2(1-\cos\theta)^2 + \frac{1}{4}\big|H_{--}\big|^2(1+\cos\theta)^2 \bigg] \bigg\},   \nonumber \\
\frac{d^2 \Gamma^{K^*}[\lambda_{\ell_i} =\frac{1}{2},\lambda_{\ell_j} =-\frac{1}{2}]}{ds d \cos\theta} &=& \frac{u(s)\big|G_{LQ}\big|^2}{2^{9} \pi^3 m^3_B s} \bigg\{
                 \Big|h_{\frac{1}{2},-\frac{1}{2}}\Big|^2 \bigg[\frac{1}{2} \big|H_{00}\big|^2 \sin^2\theta \nonumber \\
                 && + \frac{1}{4}\big|H_{++}\big|^2(1+\cos\theta)^2 + \frac{1}{4}\big|H_{--}\big|^2(1-\cos\theta)^2 \bigg] \bigg\},   \nonumber \\
\frac{d^2 \Gamma^{K^*}[\lambda_{\ell_i} =-\frac{1}{2},\lambda_{\ell_j} =-\frac{1}{2}]}{ds d \cos\theta}    &=&  \frac{u(s)\big|G_{LQ}\big|^2}{2^{9} \pi^3 m^3_B s} \bigg\{
                   \Big|h_{-\frac{1}{2},-\frac{1}{2}}\Big|^2 \bigg[\big|H_{0t}- H_{00}\cos\theta\big|^2 \nonumber \\
                 && + \frac{1}{2}\big|H_{++}\big|^2\sin^2\theta + \frac{1}{2}\big|H_{--}\big|^2\sin^2\theta \bigg] \bigg\}.
\end{eqnarray}

\end{appendix}


\end{document}